\documentclass[12pt,letterpaper]{article}

\usepackage{graphicx,array}
\usepackage[dvipsnames]{xcolor}
\definecolor{darkblue}{rgb}{0,0.1,0.5}
\definecolor{darkgreen}{rgb}{0,0.5,0.2}
\definecolor{darkred}{RGB}{153,26,0}

\definecolor{seablue}{rgb}{0,0.2,0.6}
\usepackage{latexsym}
\usepackage{amssymb, amsmath}
\usepackage{bm}        
\usepackage[numbers,sort&compress]{natbib}
\usepackage{bm,relsize}
\usepackage{slashed}
\definecolor{viola}{RGB}{134,41,198}
\usepackage{mathrsfs}
\usepackage{hyperref} 
\hypersetup{
    colorlinks=true,       
    linkcolor=darkblue,          
    citecolor=BrickRed,        
    filecolor=magenta,      
    urlcolor=MidnightBlue           
}
\usepackage[all]{hypcap} 
\usepackage{subcaption}

\usepackage[font=small,labelfont=bf,labelsep=period]{caption}

\newcommand{\GeV}{\mathrm{GeV}}

\newcommand{\Mpl}{M_{\rm Pl}}

\newcommand{\be}{\begin{equation}}
\newcommand{\ee}{\end{equation}}

 \date{\today}

\begin{document}

\begin{flushright}

\end{flushright}
\vspace{.6cm}
\begin{center}
{\LARGE \bf Jump Starting the Dark Sector\\ with a Phase Transition}\\
\bigskip\vspace{1cm}
{
\large Michele Redi, Andrea Tesi}
\\[7mm]
 {\it \small
INFN Sezione di Firenze, Via G. Sansone 1, I-50019 Sesto Fiorentino, Italy\\
Department of Physics and Astronomy, University of Florence, Italy
 }
\end{center}

\vspace{2cm}

\centerline{\bf Abstract} 
\begin{quote}
We study the possibility to populate the dark sector through a phase transition. 
We will consider secluded dark sectors made of gauge theories, Randall-Sundrum scenarios and conformally coupled elementary particles.
These sectors have in common the fact that the action is approximately Weyl invariant, implying  that particle production 
due to time dependent background is strongly suppressed. In particular no significant production takes place during inflation
allowing to avoid strong isocurvature constraints from CMB. As we will show, if the scale of inflation is large compared to the dynamical mass scale, 
these sectors automatically undergo a phase transition that in the simplest cases is controlled by the Hubble parameter. If the phase transition takes place during reheating 
or radiation the abundance obtained can be larger than particle production and production from the SM plasma. For phase transitions completing during
radiation domination, the DM mass is predicted in the range $10^8$ GeV while larger values are required for phase transitions occurring during reheating.
\end{quote}

\vfill
\noindent\line(1,0){188}
{\scriptsize{ \\ E-mail:\texttt{  \href{mailto:michele.redi@fi.infn.it}{michele.redi@fi.infn.it}, \href{andrea.tesi@fi.infn.it}{andrea.tesi@fi.infn.it}}}}
\newpage

\section{Introduction}
The possibility that Dark Matter (DM) is just gravitationally coupled to the visible sector is certainly worrisome, since it comes with no obvious experimental and observational signatures, but it is one that we have to start embracing seriously. If gravity is all we got, we should at least explain how dark sectors can ever be populated and the observed DM relic abundance reproduced. It would be remarkable if gravity itself is responsible for the DM genesis.

One unavoidable contribution to the dark sector abundance is the production through tree level graviton exchange, also known as gravitational freeze-in \cite{Garny:2015sjg,Tang:2016vch}.  This is efficient if the reheating temperature is close to the experimental bound from inflation $T_R|_{\rm max} \approx 5\times 10^{15}\,\GeV$ and rapidly declines for smaller values. Another generic mechanism is particle production due the time dependent background \cite{ford,Chung:1998zb}. This is at work during inflation for minimally coupled scalars and for massive vector fields \cite{Graham:2015rva} leading to masses as low as $10^{-5}$ eV. In the first case this however comes at the price of 
strong constraints from isocurvature perturbations \cite{Chung:2004nh}, while for the vector it relies on the Stueckelberg mechanism for mass generation \cite{Redi:2022zkt}. 

In this paper we introduce a new mechanism that allows secluded sectors to be populated even in absence of any coupling to the Standard Model (SM) and to the inflaton.
We consider interacting dark sectors with a dynamical mass scale $M$, arising either from confinement or from spontaneous symmetry breaking. 
If the scale of inflation $H_I> M$ the sector is in the unbroken phase during inflation and undergoes a phase transition during reheating 
or radiation domination. Assuming no thermal population the energy gained from the phase transition, of order $M^4$, populates the dark sector whose lightest state is automatically stable and constitutes the DM candidate. In the simplest scenarios we find,
\begin{equation}
\frac{\Omega h^2}{0.1} \approx  \frac {T_R}{10^{12}\rm GeV}  \left(\frac {M}{10^{8}\rm GeV}\right)^2 \,,~~~~~~ T_R< \sqrt{M \Mpl}
\end{equation}
leading to heavy DM scenarios.

Production from a phase transition is particularly transparent for Weyl invariant sectors because inflation automatically prepares 
the system in  a false vacuum empty state.  This is attractive as it avoids strong isocurvature constraints from inflationary production. 
The only relevant scales in the evolution are Hubble and $M$ so that the phase transition is triggered when $H\sim M$, 
a condition realized during reheating or radiation domination. This is complementary to particle production due to the time dependent background 
but as we will see typically leads to a larger abundance. The contribution can also dominate gravitational freeze-in  depending on the reheating temperature.

As examples we consider asymptotically free dark gauge theories, strongly coupled Conformal Field Theories (CFT) with deformations (and their holographic realization in Randall-Sundrum scenarios) and conformally coupled elementary scalars (with a second order instability). If the scale of inflation is large compared to the mass all these scenarios undergo a phase transition after inflation that populates the dark sector. The lightest state of the sector is automatically stable providing a natural DM candidate. Moreover, these theories are approximately Weyl invariant at high energies so that no significant particle production happens during inflation. The assumption of a Weyl invariant sector is not very restrictive being only violated by minimally coupled scalars that are naturally associated to spontaneous breaking of a global symmetry.

The paper is organized as follows. In section \ref{sec:scalar} we discuss in detail the phase transition mechanism for a conformally coupled 
scalar with an instability. In this case one can follow explicitly the dynamics of the scalar during the phase transition and determine the abundance.
We show in this simple example that the phase transition dominates particle production in the expanding universe and gravitational freeze-in if $T_R< 10^{13}$ GeV. 
In section \ref{sec:gauge-theories} we consider gauge theories arguing that the confinement phase transition can also successfully 
populate the dark sector. We also study inflationary production that is controlled by the $\beta-$functions of the theory. 
For strongly coupled CFTs we use the AdS/CFT correspondence to determine the inflationary production. A potentially sizable contribution 
exists in this case that is associated to the explicit breaking a conformal invariance of gravity. We then turn to the contribution from the phase transition
that can be determined using the dilaton effective action. We summarise in \ref{sec:conclusions}. In appendix \ref{app:A} we derive analytically the abundance of conformally coupled scalar produced in radiation domination.
 
\section{Phase Transition Mechanism}\label{sec:plot}

Approximately Weyl invariant dark sectors such as gauge theories are very difficult to produce if they are only gravitationally coupled to the SM. 
In fact in the limit of exact Weyl invariance time dependence of the background drops out from the classical equations of motions
eliminating particle production during and after inflation. Particle production is then controlled by the explicit breaking of Weyl symmetry
that eventually induces a mass scale $M$.

These sectors can however be produced gravitationally from the SM thermal bath through tree level graviton exchange (also known as, gravitational freeze-in) \cite{Garny:2015sjg,Tang:2016vch}. This type of thermal production is only efficient for very large SM reheating temperature. As we will discuss, another mechanism can be relevant due to the dynamics of the mass scale. In the case where the dark sector does not thermalize among itself, it can gain an energy density proportional to $M^4$ when Hubble becomes comparable with $M$. 

In the rest of this section we give an outline of the two contributions, writing the general scaling of the energy density produced in the two cases.

\subsection{Gravitational freeze-in: tree level graviton exchange}
After reheating of the SM to temperature $T_R$ dark sector states are produced through $s$-channel graviton exchange from the SM plasma. This production is analogous to a UV dominated freeze-in, since it originates from SM thermal initial states. In the approximation that $T_R$ is much larger than SM thresholds and than the mass $M$, it was shown in \cite{Redi:2020ffc} that the abundance of free particles can be  written in a completely general way as\footnote{We report throughout the ratio
of energy to entropy densities. The energy fraction is then given by $\Omega h^2 = 0.27/{\rm eV} \rho/s$.}
\begin{equation}\label{eq:GR-0}
\frac {\rho} s\bigg|_{\rm GR}= 6 \times 10^{-6} M c_D \left(\frac {T_R}{\Mpl}\right)^3\,, ~~~~~ T_R > M\,,
\end{equation}
where $c_D$ is the central charge ($c_D=4/3, 4 ,16$ for a conformally coupled scalar, Weyl fermion, massive gauge field respectively) of the approximate relativistic CFT that 
describes the dark sector at these energies.

If interactions allow the dark sector to thermalize in the relativistic regime, it develops a dark sector temperature $T_D$, given by $T_D= 0.25 (c_D/g_D)^{1/4}(T_R/\Mpl)^{3/4} T$, where $T$ is the SM temperature and $g_D$ is the number of degrees of freedom. In such a case the abundance becomes 
\begin{equation}\label{eq:GR-thermal}
\frac {\rho} s\bigg|_{\rm GR,\, therm.}= 1.5 \times 10^{-4} M g_D \left(\frac {c_D}{g_D}\right)^{3/4}\left(\frac {T_R}{\Mpl}\right)^{9/4}\,,
\end{equation}
that is larger than for free particles.

\subsection{Phase transition and particle-production}
Gravitational freeze-in becomes inefficient when $T_R$ is small, as it stems from the formulas, leaving a practically empty dark sector at the onset of radiation domination. In such a case, the dark sector is only characterized by the relative size of the Hubble parameter $H$ and the mass scale $M$.

We wish to argue that when $H\sim M$ an energy density of order $\sim M^4$ becomes available in the dark sector, which soon after starts to redshift as matter.
Under this assumption the abundance is found to be
\be\label{stima}
\frac{\rho}{s}\bigg|_{\rm PT}\sim M \, \mathrm{Min}\left[\left(\frac{M}{\Mpl}\right)^{\frac32}, \frac{T_R M}{\Mpl^2}\right]\,,
\ee
independently of the details of inflation.
The above formula captures the situation where critical condition $H\approx M$ happens both during radiation or during reheating. 

While the formula above is simply based on the argument that an energy density of order $ M^4$ becomes dynamical at a critical time, we will show that it  represents at least two well defined mechanisms by which a dark sector can be populated: phase transitions and particle production due to the expanding universe.
\begin{itemize}
\item \textbf{Phase Transition}. Thanks to the generation of the dynamical scale $M$ at the phase transition, the dark sector gains an energy density $\Delta V \approx M^4$. Since the UV contributions from gravitational freeze-in and inflationary production are very  suppressed, it is a good approximation to consider the phase transition to happen practically at zero temperature. In a cosmological scenario, the critical parameter in this case can be identified as $H\approx M$.
\item \textbf{Particle Production}. Thanks to the expansion of the Universe, when $H\approx M$ the Fourier modes of the quantum fields undergoing the phase transition (or simply acquiring a mass term) experience a large deviation from adiabaticity. Such condition signals the non-thermal production of non-relativistic particles that can be determined computing  Bogoliubov coefficients.
\end{itemize}

In both cases the expected scaling is the one of eq.~\eqref{stima}. However, as we will show explicitly in a few cases, the contribution from the phase transition can be enhanced at weak coupling and it can be the dominant source of energy allowing for a `jump start' of secluded dark sectors.

\section{Elementary conformal scalar}\label{sec:scalar}
We first consider a scalar field with conformal coupling to the curvature. The generic renormalizable theory is described by the lagrangian
\be\label{eq:scalar}
\mathscr{L}= \frac{(\partial_\mu \phi)^2}{2} -\frac{1}{2} \mu^2 \phi^2 +\frac{1}{12}\phi^2 R -\frac{\lambda}{4}\phi^4\,.
\ee
The coupling to curvature guarantees that classically for $\mu=0$ the action is invariant under a Weyl transformations, $g_{\mu\nu} \to \Omega(x)^2 g_{\mu\nu}$ and $\phi\to \Omega(x)^{-1} \phi$ so that the scale factor can be removed from the equations of motion. A related important fact is that the coupling to curvature generates a positive mass squared $2H_I^2$ during inflation. Weyl invariance is explicitly broken by the mass term and the running of the couplings (see for example \cite{collins}).

We assume that the sector is  decoupled from the SM, so that it can be populated only through gravitational effects.
In this work, we wish to distinguish between quantum production due to the non-adiabatic evolution of the Bunch-Davies vacuum, and production from a phase transition. A second order phase transition is possible if $\mu^2=-M^2/2$, while $\mu^2=M^2$ is the other possible branch. This parametrization is chosen to denote with $M$ the mass of the scalar
both in unbroken and broken phase.

\subsection{Particle production from time dependent background}
 \begin{figure}
    \centering
    \includegraphics[width=0.65\linewidth]{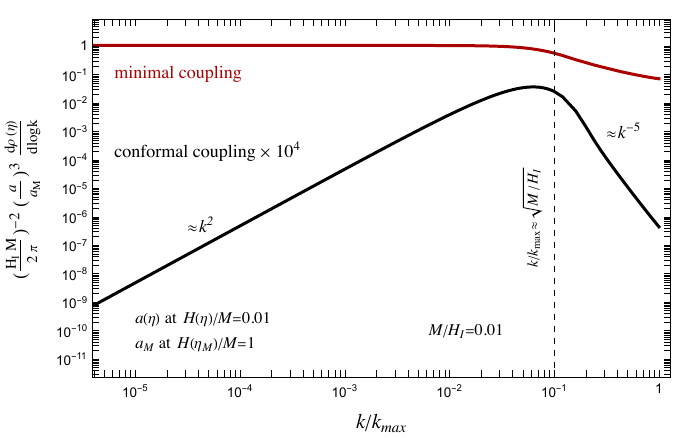}
  \caption{\it Energy density per unit logarithmic mode $a^4 d\rho/d\log k$ of eq.~\eqref{eq:density}, normalized to $(H_I M/2\pi)^2 a a_M^3$. The plot shows the energy of real scalars of mass $M$ conformally (black)  and minimallly (red) coupled. The spectrum is for $M/H_I=0.01$ and it is cut-off at $k_{\rm max}= H_I a_e$.}
    \label{fig:power}
\end{figure}

We start reviewing particle production in the scenario with no interactions and positive mass term $M^2=\mu^2$. 
In a expanding universe particle are produced due to the time dependence of the background.
This is controlled by the explicit breaking of Weyl invariance that for the conformally coupled scalar is the mass term.  

Upon rescaling the field by the scale factor $a$, $\phi= v/a$, the equation of motion in conformal time takes the form
\begin{equation}
v_k''(\eta) +\omega_k^2(\eta) v_k(\eta)=0\,, ~~~~~~~\omega_k^2(\eta) = k^2 + M^2 a^2(\eta)
\label{eq:scalar-conformal}
\end{equation}
where $M$ is the physical mass today. 

As reviewed in the appendix the energy density per unit logarithmic interval produced by the non-adiabatic evolution of eq.~\eqref{eq:scalar-conformal} can be written as
\be\label{eq:density}
a^4 \frac{d \rho}{d\log k}= \frac{k^3}{2\pi^2}  \lim_{\eta\to \infty}\left[\frac{|\partial_\eta v_k|^2}{2}+\frac{\omega_k^2|v_k|^2}{2}-\frac{\omega_k}{2}\right]= \frac{k^3}{2\pi^2} \lim_{\eta\to \infty}\omega_k |\beta_k|^2\,,
\ee
where $\beta_k$ are the Bogoliubov coefficients associated to the transformation from the vacuum at early times to the vacuum infinity.
In the above formula $v$ is the wave function associated to the quantum operator of the field $a\phi$ that annihilates the Bunch-Davies vacuum and therefore it has initial conditions in the asymptotic past as $v_k(-\infty)=e^{-i k\eta}/\sqrt{2k}$. 

Particle production is most efficient when there is a large deviation from adiabaticity. In this case the distortion from the initial plane wave solution happens when the mass becomes relevant, i.e. $k/a\approx M$. The abundance, due to the overall $1/a^4$ factor, will  be dominated by
\be \label{eq:peak}
 H=M=\frac k  a\,,
 \ee
where the first condition is selected as otherwise the density is suppressed by the extra redshift. 
The first condition is realized in radiation domination for ,
\begin{equation}
T_R> T_* \equiv 0.55\left(\frac{100}{g_*}\right)^{1/4} \sqrt{M \Mpl}
\end{equation}
and otherwise during reheating. We will take $g_*=100$ in what follows. Therefore the production will be peaked at scales shorter than cosmological, corresponding to co-moving wave-numbers
\begin{equation}
k_{\rm peak} =\displaystyle
\left\{\begin{array}{cc} \displaystyle a_R \sqrt{H_R M} &~~~~~~~T_R> T_* \\
\displaystyle a_R \sqrt{H_R M} \left(\frac {H_R}M\right)^{1/3}& ~~~~~~~T_R< T_*\end{array}\right.
\end{equation}
respectively during radiation domination or reheating.

The amount of energy density stored in the scalar field from particle production only depends on the scale $M$, since it is the only scale of the problem. From eq. (\ref{eq:peak}) the energy density when $H=M$ can only be a function $\rho|_{H=M} =\kappa M^4$, where $\kappa$ is a coefficient that can be calculated precisely from eq.~\eqref{eq:density}. In terms of $\kappa$, the yield of scalars at late times is 
\begin{equation}
\frac {\rho}s =\kappa \left\{\begin{array}{cc} \displaystyle 0.14 \, M\left(\frac{M}{\Mpl}\right)^{\frac32} &~~~~~~~T_R \gtrsim T_*\\ &\\
\displaystyle \frac{M^4}{s(T_R)}\left(\frac {a_M}{a_R}\right)^3=  T_R \frac{M^2}{4\Mpl^2} &~~~~~~~T_R\lesssim T_*\end{array}\right.
\label{eq:radvsreh}
\end{equation}
 where we have used $s=2.3 g_*^{1/4} (H \Mpl)^{3/2}$ valid in radiation domination. 
In the second equation  $a_M$ is the scale factor when $H=M$, we approximated reheating as a phase of matter domination so that $H=H_R (a_R/a)^{3/2}$.\footnote{We do not consider here the model dependent contribution to particle production from inflaton scatterings \cite{Ema:2015dka,Schiappacasse:2016nei,Ema:2018ucl,Chung:2018ayg,Mambrini:2021zpp,Basso:2022tpd,Kaneta:2022gug}.} 

Numerical computations of the abundance of conformally coupled real scalars can be found in \cite{Ema:2018ucl,Ema:2019yrd} using Bogoliubov coefficients. In figure  \ref{fig:power} we show the energy density $a^4 d\rho/d\log k$ evaluated at a conformal time corresponding to $H<M$.

However, here we would like to present analytical results, decomposing the contribution in different epochs, see also \cite{Herring:2019hbe,Herring:2020cah}. We start with radiation domination where the contribution is the largest, and then  discuss production during reheating and the subdominant production during inflation (where Weyl invariance is a good symmetry).

\paragraph{Radiation}~\\
We first consider the case where the condition (\ref{eq:peak})  takes place during radiation, ie. $T_R\gtrsim \sqrt{M \Mpl}$.
During radiation where the scale factor reads $a= a_R (1+ a_R H_R \eta)$ the equation of motion (\ref{eq:scalar-conformal}) can be solved explicitly in terms 
of parabolic cylinder functions. As shown in the appendix  this allows to analytically determine the Bogoliubov coefficients as 
\be
|\beta(z)|^2=\frac{e^{-\frac{3 z}{4}} \left(e^z+1\right) }{4\sqrt{\pi z}} \left(\frac{z}{4\pi}  \bigg|\Gamma \left(\frac14 + i\frac{z}{4\pi}\right)\bigg|^2+  \bigg|\Gamma \left(\frac{3}{4} +i \frac{z}{4\pi}\right)\bigg|^2\right)-\frac12
\,,\quad z\equiv \frac {\pi k^2}{a_R^2 H_R M}
\label{eq:bogo}
\ee
From eq. (\ref{eq:bogo}) the energy density per unit logarithmic momentum interval is given by,
\be
a^3 \frac{d\rho}{d\log k}= \frac{k^3}{2\pi^2} M \left\{\begin{array}{cc} \displaystyle \frac{\Gamma[3/4]^2}{2\pi}\, a_R \frac{\sqrt{H_R M}}{k}
& k\ll a_R \sqrt{H_R M} \\
\displaystyle a_R^8\frac{M^4 H_R^4}{64 k^8}& k\gg a_R \sqrt{H_R M} \end{array}\right.\,.
\label{eq:rho-cft-scalar}
\ee
Asymptotic behaviors at small and large $k$ agree with other results in the literature (see for example \cite{Herring:2020cah}).
Analytically it is possible to compute the position of the peak of the power spectrum at $k_{\rm peak}=0.629279 a_R \sqrt{M H_R}$
where $a^3 \rho_{\rm peak}=0.00092\, a_R^3 M^{5/2}H_R^{3/2}$. Performing the integral over momentum the total abundance is given by
\be
\frac{\rho}{s}\bigg|_{\rm quantum}= \frac{\mathcal{C}}{2\pi^{\frac72}}\,  \frac{M^{5/2}H_R^{3/2}}{s(T_R)}=0.0002\, M\left(\frac{M}{\Mpl}\right)^{\frac32}
\label{eq:quantumrad}
\ee
where we used $\mathcal{C}= 0.164341$, see appendix. As anticipated, the DM abundance is only 
set by the mass of the scalar, for which we get the prediction $M_{\rm DM}=5.7\times 10^8\, \GeV$.

\paragraph{Reheating}~\\
If the scalar is produced during reheating ($T_R\lesssim \sqrt{M \Mpl}$) one should solve the wave-equation in matter domination.
The analytic solution is given in term of Heun functions. To estimate the final abundance we can however simply rescale 
the result during radiation,
\be
\frac{\rho}{s}\bigg|_{\rm quantum}\approx \frac{\mathcal{C}}{2\pi^{\frac72}}\,  \frac{M^{5/2}H_*^{3/2}}{s(T_R)} \left(\frac {a_M}{a_R}\right)^3\approx 10^{-4}  \, T_R \frac{M^2}{\Mpl^2}\,.
\ee

 \begin{figure}[t]
    \centering
        \includegraphics[width=0.47\linewidth]{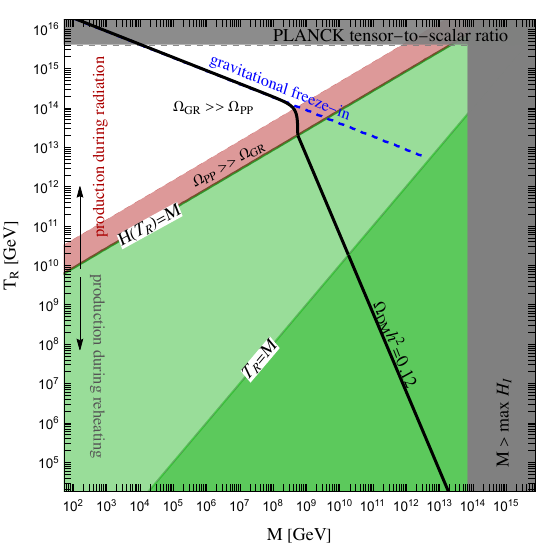}~
        \includegraphics[width=0.47\linewidth]{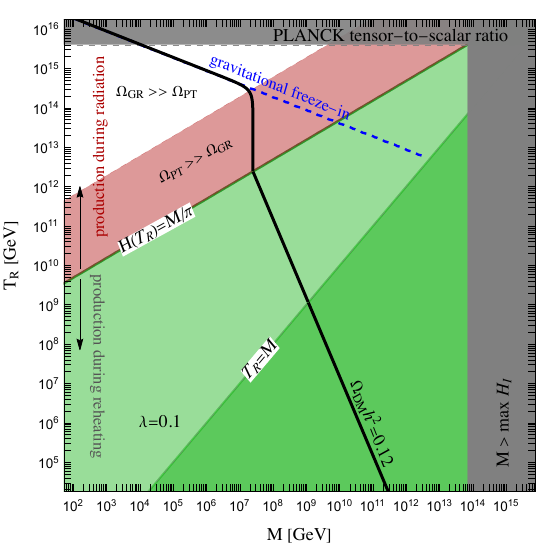}
  \caption{ \label{fig:conformal-scalar} \small \it Parameter space of the conformally coupled scalar. The DM abundance is reproduced along the black solid line. The darker green region corresponds to $T_R\leq M$ where the gravitational freeze-in (blue line) becomes inefficient (see eq.~\eqref{eq:GR-0}). The lighter green region corresponds to $H_R\leq M$ where the particle production (left panel) or phase transition (right panel) happen during reheating. The red region is highlighted to show where the contribution from particle production (left panel) or phase transition (right panel) generated in radiation domination dominate over the gravitational freeze-in. Gray regions are excluded.}
\end{figure}

In figure \ref{fig:conformal-scalar} we show the regions where production from the time dependent background dominates the gravitational 
contribution. This is only realized in a small region for $T_R> T_*$ while it is generic in the opposite regime, i.e. if DM is produced during reheating.

\paragraph{Inflation}~\\
Finally we discuss what happens during inflation, in order to show that Weyl symmetry forbids any sizeable particle production in this epoch, as anticipated in the introduction and section \ref{sec:plot}. This derivation will be particularly useful in the next sections. The equation of motion in de Sitter reads,
\begin{equation}
v_k'' + k^2 v + \frac {M^2}{H_I^2 \eta^2} v_k=0\,.
\label{eq:scalarinf}
\end{equation}
To leading order in $M^2/H_I^2$ the solution of the equation above with Bunch-Davies boundary conditions is given by
\be
v_k=\frac{e^{-ik\eta}}{\sqrt{2k}} - i \frac{M^2}{H_I^2} \frac{e^{ik\eta}}{\sqrt{2k}} \frac{\big[\pi+i \mathrm{Ei}(-2k\eta)]}{2} +O(\frac{M^4}{H_I^4})\,,
\ee
where  $\mathrm{Ei}(z)$ is the exponential integral function. The solution above is valid for $k \eta< 1$ since modes are produced when $k/a = H_I$ 
and could be expanded in $\log$ terms.
To determine the energy density at a later time one should evolve this solution during reheating and radiation.
For the incoming plane wave we have already performed this computation above, leading to a production peaked when $H=M =k/a$. 
For the second term proportional to the outgoing wave to leading order in $M^2$  the evolution is trivial so that the Bogoliubov coefficient can 
be directly computed from the wave-function at the end of inflation. The energy density at later times is thus
\begin{equation}
\frac {d\rho}{d\log k} \sim \frac {k^3}{2\pi^2} \frac{\sqrt{k^2+M^2 a^2}}{a^4} \frac {M^4}{H_I^4}\,.
\label{eq:infscalar}
\end{equation}
From this formula it follows that the energy density from inflationary fluctuation is dominated by the highest
momentum modes $k= a_e H_I$ that are produced at the end of inflation and  renter the horizon right after with an energy density $M^4/(2\pi^2)$.
Comparing with eq.~(\ref{eq:rho-cft-scalar}) we can see that this amount of energy is always negligible.

\subsection{Production from phase transition}
\label{sec:PT}

When $\mu^2<0$ the origin of the potential is unstable in flat space. The minimum of the potential is at $\phi= \mu/\sqrt{\lambda}$
where the mass is $M^2=- 2 \mu^2$ and the $Z_2$ symmetry is spontaneously broken\footnote{The model as it stands suffers from domain wall problems due to the spontaneously broken $Z_2$ symmetry. This can be easily solved by adding a small cubic term that makes domain walls unstable. The conformal scalar illustrates the general fact that production from the phase transition dominates over quantum production.}. During inflation however the coupling to curvature produces a positive mass $2 H_I^2$ that stabilizes the  field at the origin if $H_I>M/2$.\footnote{ A similar mechanism can be found in Ref. \cite{Babichev:2020xeg} where however the $Z_2$ symmetry is restored during the expansion of the universe.} Even if the scalar is initially displaced after few e-foldings of inflation the field reaches $\phi=0$, see \cite{Redi:2022zkt}.

After inflation the effective mass drops during reheating and vanishes in radiation since $R=0$.
As a consequence the origin of the potential becomes unstable and  the field starts to evolve towards the minimum
through a second order phase transition. 

The inflationary production can be studied as in the previous section.
During inflation fluctuations of the field satisfy eq.~(\ref{eq:scalarinf}) with a negative squared mass. To leading order this however  leads to exactly the same
amount of relativistic modes since this is proportional to $M^4/H_I^4$. Therefore also in this case the production during inflation is negligible.
This shows explicitly that the phase transition occurs in an empty dark sector. 

The dark sector is thus populated only through gravitational freeze-in from the SM thermal bath and by the latent heat associated to the phase transition. 

\paragraph{Abundance}~\\
To determine the abundance we need to estimate the time when the phase transition completes.
The time scale for the second order PT is set by the mass $M^2$ of the field.  We assume that the scalar field starts to oscillate as non-relativistic matter after a time $t_*=2\pi/M$. Since $H\sim 1/t$ we use as criterium that the phase transition completes for  $H_*=M/(2\pi)$. 
With this criterium the DM abundance can be determined from the latent heat $\Delta V=M^4/(16\lambda)$ assuming that the energy redshifts as matter after the phase transition,
\be
\frac{\rho_\phi}{s}\bigg|_{\rm PT}=\frac{\Delta V}{s(T_*)} \approx \frac{M}{\lambda}   \mathrm{min}\left[0.15\frac {M^{3/2}}{\Mpl^{3/2}}\,,0.4 \frac{M T_R}{\Mpl^2} \right]\,.
\label{eq:abPT1}
\ee
If the relaxation time of the phase transition is longer, a larger abundance is produced. Therefore the estimates above can be considered as a lower bound.  
This provides an explicit computation of the contribution discussed in eq.~\eqref{stima}.

Note that the abundance from the phase transition has the same dependence on mass and reheating temperature as the contribution (\ref{eq:quantumrad}).
This follows from the fact that in both cases DM is produced when $H\sim M$. The coefficient is larger and enhanced for small $\lambda$ so that the abundance obtained for equal mass is always larger than the free theory. 

The comparison with the quantum production and tree-level gravitational production is shown in figure \ref{fig:conformal-scalar}.\footnote{We assume here that no thermalization takes place in the relativistic regime. When this happens the PT takes place when $T_D^*\sim \mu/\sqrt{\lambda}$ and the abundance is roughly $\rho/s\sim \xi^3 T_D^*$.} Gravitational freeze-in dominates when the mass $T_R> \sqrt{M \Mpl}$ except for the red region where the effects from quantum and phase transition are important during radiation. When reheating is not instantaneous (green region), gravitational freeze-in is subdominant and the production is dominated by the quantum production or phase transition happening during reheating.

\section{Gauge theories}\label{sec:gauge-theories}

We now turn to confining gauge theories. In the massless limit the action of fermions and gauge fields is automatically Weyl invariant at the classical level. 
Weyl invariance is violated by fermions mass terms and by quantum corrections that introduce a dynamical scale $\Lambda$ in the theory that controls the mass of the all the hadrons in the chiral limit. Weyl symmetry becomes completely broken when the theory confines through a first or second order phase transition. We argue that the production of such sector can be analogous  to the one of the conformally coupled  scalar that undergoes a phase transition when $H\sim M$. 
Contrary to scalars gauge theories would not suffer from domain wall problems but other topological defects might exist \cite{Yamada:2022imq}. 

To start with we consider the production of this sector during inflation, due to the violation of Weyl invariance from the running of gauge couplings.
Using the results of the conformal scalar this contribution is negligible, see also \cite{Benevides:2018mwx} for related work. 
The dark sector can instead be populated by the confinement phase transition leading to  DM mass around $10^8$ GeV.

\subsection{Quantum production: suppressed by $\beta$-functions}
\label{sec:weyl-anomaly}

In asymptotically free gauge theories, at energies larger than the confinement scale $\Lambda$, the theory can be effectively described as a weakly coupled CFT with marginal deformations.  The deviation from conformality in the chiral limit arises at the quantum level, conformal and Weyl invariance are violated by the running of the gauge coupling that eventually leads to confinement. 

In the perturbative regime the quantum production in an expanding universe can be computed as  in \cite{Dolgov:1981nw,Dolgov:1993vg}. 
For SU($N$) gauge theories with $N_F$ flavors the 1-loop $\beta-$function of the gauge coupling is
\be\label{eq:running}
\beta(g) = -\frac{g^3}{16\pi^2}b_0\,, \quad\quad b_0=\frac{11}{3}N-\frac{2}{3}N_F\,.
\ee
The 1PI effective action in flat space has thus the form
\be
\mathscr{L}=-\frac{1}{4g^2}\left[1+\frac{g^2 b_0}{16\pi^2} \log (\frac{-\square}{M^2}) \right] G_{\mu\nu}^2  + \cdots 
\ee
where $g$ is the gauge coupling renormalized at the scale $M$. Weyl invariance can be formally restored promoting $M$ to a field transforming as $M\to \Omega^{-1} M$.
This implies that in conformally flat background $g_{\mu\nu}=e^{2\Omega}\eta_{\mu\nu}$ the quantum effective action acquires a new term that depends on the scale factor,
\be \label{eq:rescaled}
\mathscr{L}=-\frac{1}{4g^2}\left[1+\frac{g^2 b_0}{16\pi^2} \log (\frac{-\square}{M^2}) \right] G_{\mu\nu}^2 -\frac{1}{4} \left(-\frac{b_0}{8\pi^2}\right)\Omega\, G_{\mu\nu}^2 +\cdots\,.
\ee
For $\Omega = \log a$ the equation of motion (in conformal coordinates) are then easily derived from the above effective action
\be
\frac{1}{g^2(-\square)}\partial_\mu G^{\mu\nu} - G^{\mu\nu}\frac{b_0}{8\pi^2} \partial_\mu \log a =0\,,
\label{eq:anomaly}
\ee
The above equation is better rewritten in Fourier space, making explicit the derivative with respect to conformal time and neglecting the $\log(a)$ as compared to its derivative \cite{Dolgov:1981nw}. As expected only the two transverse polarizations $\vec A_T$  have non-trivial dynamics while the other two $A_\eta$ and $\vec A_L\sim \vec k$ are non-dynamical. The equation for the transverse components is
\be
A_T'' +k^2 A_T + \frac{b_0 g^2}{8\pi^2} \, \frac{a'}{a} A_T'=A_T'' +k^2 A_T -2 \Delta \, \frac{a'}{a} A_T'=0\,, \quad \Delta =-\frac{b_0 g^2}{16\pi^2}\,,
\ee
where $\Delta$ coincides with the anomalous dimension of $G_{\mu\nu}^2$.\footnote{We find a factor of 2 difference with respect to the derivation in \cite{Dolgov:1981nw}.} We can eliminate the first derivative with the rescaling $A_T \to a^{\Delta} v.$ In terms of the new mode function $v$, the equation has the same form of a conformal scalar eq.~\eqref{eq:scalar-conformal}, with frequency given now by
\be
\omega_k^2=k^2+\Delta (\Delta+1) \frac {a'^2}{a^2}-\Delta \frac {a''}a \,.
\ee
The energy density of the transverse mode can thus be computed using eq.~\eqref{eq:density}. During inflation the frequency is
\be
\omega_k^2\big|_{\rm dS}= k^2+\frac{\Delta}{\eta^2}\,.
\ee 
Therefore the equation is identical to the one of the conformal scalar during inflation with the replacement $M^2/H_I^2 \to \Delta$.
Since confining gauge theories have negative anomalous dimension $\omega_k^2$ becomes negative for small $k$ but as we discuss the instability is irrelevant. 
Since the gauge fields are massless we can directly borrow the result for the scalar (\ref{eq:infscalar}) so that the energy density of relativistic gluons is given by,
\begin{equation}
\rho_{\rm gluons} \sim (N^2-1) \frac{H_I^4}{\pi^2}\Delta^2 \frac {a_e^4}{a^4}\ll H^4
\end{equation}
Similar results can be found in \cite{Benevides:2018mwx}.
This contribution is again negligible as we will see.

During radiation the frequency becomes
\be
\omega_k^2\big|_{\rm RD}=k^2-\frac{\Delta}{\eta^2}\,,
\ee
so that particle production  is even more suppressed in this phase (in the  perturbative regime)  since $\eta$ grows.

It is useful to notice that the deviation from Weyl invariance is proportional to  the anomalous dimension of the operator $G_{\mu\nu}^2$.
This allows  to generalize these results to theories with scalars and fermions. Due to the wave-functions renormalization the classical 
equations are corrected by a term proportional the anomalous dimension of the kinetic term. One thus finds that the quantum production
of particles is controlled by the square of the anomalous dimension. 

\subsection{Phase transition: Glueball dark matter}

In analogy with the conformal scalar we now consider the population of the dark sector through the phase transition.  
For simplicity we focus on  pure glue gauge theories that give rise to glueballs in the confined phase. 
The lightest glueball in particular is accidentally stable and provides an excellent DM candidate, see \cite{Cline:2013zca,Boddy:2014yra,Soni:2016gzf,Forestell:2017wov,Acharya:2017szw,Jo:2020ggs,Redi:2020ffc,Gross:2020zam}. 
A wealth of lattice results are in particular available for SU(N) gauge theories. For $N=3$ in particular latent heat and lightest CP even glueball are found \cite{Beinlich:1996xg,Morningstar:1999rf,Borsanyi:2012ve},
\begin{equation}
L_h = 1.4 \Lambda^4 \,,~~~~~~~~~~~~~~~~M_{0^{++}}= 5.5 \Lambda
\label{eq:lattice}
\end{equation}
where $\Lambda$ is the critical temperature of the finite temperature de-confinement phase transition.

For $H_I> \Lambda$ the gluons are free during inflation. There are several ways to argue for this. 
First de Sitter space can be roughly treated as a thermal bath of temperature  $T_{\rm ds}=H_I/(2\pi)$ so that for $H_I>\Lambda$ deconfinement takes place. 
A related argument is that in euclidean signature dS is 4-sphere of radius $1/H_I$ so that the space is smaller than the size of the hadrons.
In real time the hadrons would have a size larger than the horizon so there is no local physics can sustain them.
After inflation, Hubble drops during reheating and then radiation so that a phase transition to the confined vacuum 
will take place to the confined vacuum. If the  system has not thermalized in the relativistic regime as will be the case for large masses considered here,
the phase transition happens out of equilibrium. Nevertheless given that the only scale available is Hubble we expect the phase transition to happen when $H\sim \Lambda$.
Indeed $1/\Lambda$ is also  the typical time scale on interactions.

\paragraph{Abundance}~\\
Given the latent heat in (\ref{eq:lattice}) for the thermal phase transition we then expect an energy $\Lambda^4$ to be released in the dark sector
when $H\sim \Lambda$. Whereas in the context of the conformal scalar with an instability it is clear the role of $H$ in the evolution of $\phi$, here it is less evident how explicitly $H$ affects the evolution of the order parameter of Yang Mills towards confinement. It is tempting to interpret $H$ as a control parameter for a potential at finite three-dimensional volume, rather than finite temperature.

Assuming an energy $\Lambda^4$ to be released for $H=\Lambda$, we can estimate the  abundance of glueballs as in eq. (\ref{eq:abPT1}),
\be
\frac{\rho_{\rm DG}}{s}=\frac{\Delta V}{s(T_R)}\frac {a_\Lambda^3}{a_R^3} \sim 0.1 \, \Lambda\, {\rm Min}\left[\left(\frac {\Lambda}{\Mpl}\right)^{3/2}\,,\frac {\Lambda T_R}{\Mpl^2}\right]\,,
\label{eq:abDG2}
\ee
This rough estimate indicates a DM mass $\approx 10^8$ GeV or heavier.\footnote{The large value of the mass raises the question of the cosmological stability of glueballs. 
Even assuming negligible couplings to the SM fields the lightest glueball that is CP even can at least decay to gravitons \cite{Soni:2016gzf}. 
Naive dimensional analysis and large N counting indicate a decay rate $\Gamma \sim N^2/(128 \pi^3) M^5/\Mpl^4$
so that cosmological stability would require at least $M< 10^7$ GeV. We see two possible way out. First, the abundance could be larger 
if the relaxation time is longer than our conservative estimate $1/\Lambda$. 
Second,  as argued in \cite{Gross:2020zam} heavier glueballs (for example the lightest CP odd glueball) might be more stable 
due to accidental symmetries. If these are also populated an abundance is produced compatibly with cosmological stability.}
The dark glueballs scenario also predicts the existence of 
cosmic strings with tension $\sqrt{\mu} \sim \Lambda$ that could be within the reach of future gravity wave experiments \cite{Yamada:2022aax}.

Let us also note that as we approach confinement the effective gauge coupling in (\ref{eq:anomaly}) becomes large and 
the anomalous dimension of $G_{\mu\nu}^2$ becomes of order 1. We can then interpret the non-adiabaticity in the equation as a source 
of particle production. 

\begin{figure}[t]
    \centering
   \includegraphics[width=0.55\linewidth]{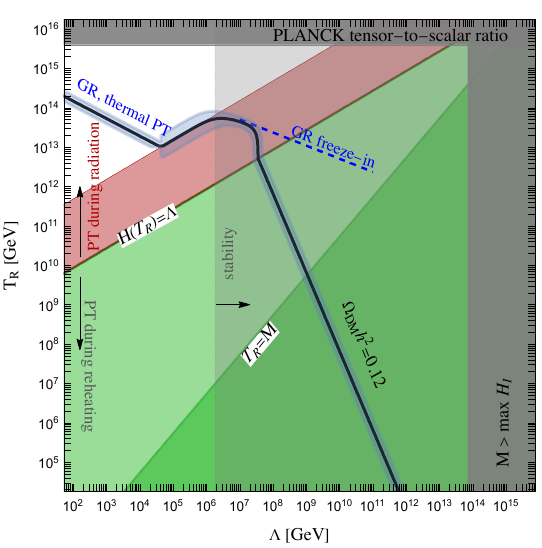}
    \caption{    \label{fig:infvsGR_DG} \small \it Parameter space of Glueball DM. The DM abundance is reproduced along the black solid line. The darker green region corresponds to $T_R\leq M_{0^{++}}$ where the gravitational freeze-in (blue line) becomes inefficient (see eq.~\eqref{eq:GR-0}). The lighter green region corresponds to $H_R\leq \Lambda$ where the phase transition happens during reheating. The red region is highlighted to show where the contribution from the phase transition generated in radiation domination dominates over the gravitational freeze-in. The light shaded region corresponds to a cosmologically unstable DM according to naive dimensional analysis estimates.}
    \label{fig:infvsGR_DG}
\end{figure}

The contribution above must be compared with the unavoidable gravitational freeze-in production, see \cite{Redi:2020ffc}. 
As shown in that reference for large reheating temperatures gluons thermalize in the relativistic regime. 
When this happens the contribution from the phase transition is already included in the finite temperature phase transition.
Thermal equilibrium however changes the abundance because the phase transition to the massive theory takes place when the dark sector temperature $T_D=\Lambda$ 
rather than $H\sim \Lambda$, enhancing the abundance. One finds for the abundance of the lightest stable glueball \cite{Redi:2020ffc}
\begin{equation}
\left(\frac{\rho_{\rm DG}}{s}\right)_{\rm GR,\, therm.}\approx 0.01 \Lambda \left(\frac {T_R}{M_{\rm Pl}}\right)^{9/4} ,
\label{eq:GRglueballs}
\end{equation}
allowing to reproduce the DM for masses as low as GeV for the largest reheating temperatures. 
Numerically the thermalization is compatible with a cosmologically stable lightest glueball for $T_R\gtrsim 10^{13}\GeV$  \cite{Redi:2020ffc}.
For lower reheating temperatures the gluons would not thermalize in the relativistic regime and confinement would occur 
out equilibrium as discussed above. After confinement, if $T_R>  M_{0^{++}}$ gravitational freeze-in would produce energetic gluons in 
the confined vacuum that would give rise to gluon jets, contributing with an abundance estimated in eq.~\eqref{eq:GR-0}. Note that this criterium is slightly different from the one employed in \cite{Redi:2020ffc}.

In figure \ref{fig:infvsGR_DG} we show the parameter space of the model. We identify three regions. The first (in lighter and darker green) where the phase transition happens during reheating and dominates the abundance, and for sufficiently small reheating temperatures it is the only contribution. The second  (in red) where phase transition is a sizable contribution even if it happens in radiation domination although production from gravitational freeze-in of dark gluons (that hadronizes in the vacuum) starts to be competitive. And the third (in white) where DM is reproduced via a phase transition at finite temperature, since the dark gluons have thermalized while relativistic (their energy is far more dominant). 

\section{Strongly coupled scenarios}

We now generalize the previous discussion to general CFTs with deformations that create a mass gap.
As discussed \cite{Redi:2020ffc} the lightest state is accidentally stable and provides an excellent DM candidate analogous to glueballs.

In general an interacting CFT is expected to have a traceless energy momentum tensor $T_\mu^\mu\equiv0$.
This implies that when coupling to a curved background the system  is automatically Weyl invariant (modulo Weyl anomaly at quantum level that does not play a role here). 
Many of the statements above thus apply in this more general setting. 
In particular the production during inflation is controlled by possible relevant or irrelevant deformations
and thus strongly suppressed as long as $H_I\gg \Lambda$, where $\Lambda$ is the dynamical scale of the CFT.
If the system however undergoes a phase transition after inflation this can populate the dark sector.

At strong coupling and large $N$ the dynamics of CFTs can be studied through the AdS/CFT correspondence
that relates the system to a weakly coupled gravitational theory in 5 dimensional Anti-de-Sitter space \cite{maldacena}.
Let us first discuss the exact (with no deformations) CFT during inflation, i.e. the CFT in a de-Sitter background.
The dual description is a gravitational theory with AdS$_5$ metric with 4D de Sitter slicing. 
The metric can be parametrized as
\be
ds^2_5= ds_4^2 f(y)^2- dy^2\,, \quad ds_4^2=\frac {1} {H_I^2 \eta^2} (d\eta^2-dx^2)\,,\quad f(y)=e^{-k_5 y}-\frac {H_I^2} {4 k_5^2} e^{k_5 y},
\ee
where $k_5$ is the five-dimensional curvature of the space while  and $H_I\ll k_5$ is the 4D Hubble expansion.

To make contact with phenomenology we wish to make 4D gravity dynamical and add SM fields external to the CFT.
To do so we introduce a UV brane that ends the space at $y=0$ and add there SM degrees of freedom. 
This gives rise to a Randall-Sundrum like scenario \cite{Randall:1999vf} where the elementary SM fields are coupled gravitationally to a CFT (see \cite{ArkaniHamed:2000ds,Hebecker_2001} for the holographic interpretation).
The presence of the UV brane corresponds to a UV cut-off  $k_5$ for the CFT. One consequence is the contribution 
to the 4D Planck mass,
\begin{equation}
\Delta M_p^2\approx  \frac{M_5^3}{k}\equiv N_{\rm eff}^2 k_5^2\,.
\end{equation}
Since we are not interested in explaining the hierarchy problem $M_5$ and $k$ will be free parameters
and the 4D Planck mass is obtained introducing localized kinetic term on the UV brane.

Note that the 5D metric has a horizon at finite distance,
\begin{equation}
y_{hor}=\frac 1 {k_5} \log \frac {k_5} {H_I}\,.
\label{eq:hor}
\end{equation}
Due to horizon an IR brane beyond $y_{hor}$ should be replaced by the horizon. This is similar to finite 
temperature where the IR brane disappears for  $T> \Lambda_{IR}$ and is replaced by the 5D black hole horizon \cite{Creminelli:2001th}.
In this case instead the black hole is replaced by a cosmological horizon. This will be important when deformations
are included that generate a mass gap.

\subsection{Inflationary production}

The construction above allows to compute the inflationary production of the CFT through the holographic dual. 
Compared to glueballs and scalars a new effect arises associated to the explicit breaking of Weyl invariance of the UV brane.
Decomposing the 5D metric in 4D gives rise to a tower of massive spin-2 fields. These are not conformally coupled and 
therefore will be produced during inflation. We can interpret this effect as the breaking of Weyl symmetry induced by
gauging 4D gravity.

Explicitly the Kaluza-Klein reduction of the 5D metric in the setup above has been performed in \cite{Garriga:1999bq}. 
One finds a massless zero mode that describes the 4D graviton and a continuous tower of massive gravitons with $M>3/2 H_I$
(corresponding to the normal brunch of de Sitter spin-2 representations).
We should then consider the production of massive gravitons in de-Sitter space. 
Since their mass is larger the dS temperature $T_{\rm dS}=H_I/(2\pi)$ they are expected to be non-relativistic and their  abundance  thermal for $M\gg T_{\rm dS}$. 
Moreover since the mass is large the five polarization of the massive graviton are expected to behave as 5 minimally coupled scalars.
The production of massive scalars in de Sitter has been studied in several papers, the abundance of scalar fields with $M>3/2 H$ produced during inflation can be found in \cite{Li:2019ves}.  The number density at the end of inflation is found to be
\begin{equation}
n(a_e;\mu)= \frac{2\pi}3 \frac{H_I^3 \mu^3}{e^{2\pi \mu}-1}\,,~~~~~~\mu=\sqrt{\frac{M^2} {H_I^2}-\frac 9 4}\,,
\end{equation} 
where $H_I$ is Hubble at the end of inflation. Note that this corresponds to less that 1 particle per Hubble volume.

In holographic theories the number density at the end of inflation can thus be obtained integrating over the whole tower of Kaluza-Klein particles weighted by the density of states $dN/dM\equiv \sigma(M)$,
\be
n_{\rm RS}(a_e)=5 \int_0^\infty dM\, \sigma(M)\, n(a_e;\mu)\,.
\ee 
The continuum density of Kaluza-Klein states starts at $M=\frac32 H$ and it is flat afterwards (neglecting corrections $H/M$), therefore we approximate it as $\sigma(M)=k^{-1}\theta(M-(3/2) H)$. This gives
\begin{equation}
n_{\rm RS}(a_e)= 0.015\frac{H_I^4}{k_5}\,.
\end{equation}
The energy density at the end of inflation is accordingly
\begin{equation}
\rho_{\rm RS}(a_e)=5 \int_0^\infty dM\, \sigma(M)\, M\, n(a_e;\mu) =0.025 \frac{H_I^5}{k_5}\,.
\end{equation}
We see explicitly in the limit $k_5\to \infty$ (CFT with no cut-off) the energy density goes to zero and no production occurs during inflation
as expected due to Weyl invariance. 

The CFT states are unstable to decay to the SM. 
However due to interactions the dark sector can thermalize or the  states with large invariant mass decay to ligher and more stable ones (see for example the discussion in \cite{Redi:2021ipn}). Assuming that this works efficiently energy density redshifts relativistically after it thermalizes due to interactions in the CFT itself . Therefore after inflation the density at reheating reads
\be
\rho_{\rm RS}\approx 0.12 \, \frac{H_I}{k_5} \left(\frac{H_I}{\Mpl}\right)^{4/3}\left(\frac{T_R}{\Mpl}\right)^{4/3} T^4\,.
\ee
This amount of energy can be sizable and even larger than the one from gravitational freeze-in for slow reheating if $k_5$ is comparable 
to $H_I$.

\begin{figure}[t]
\centering
\includegraphics[width=0.55\linewidth]{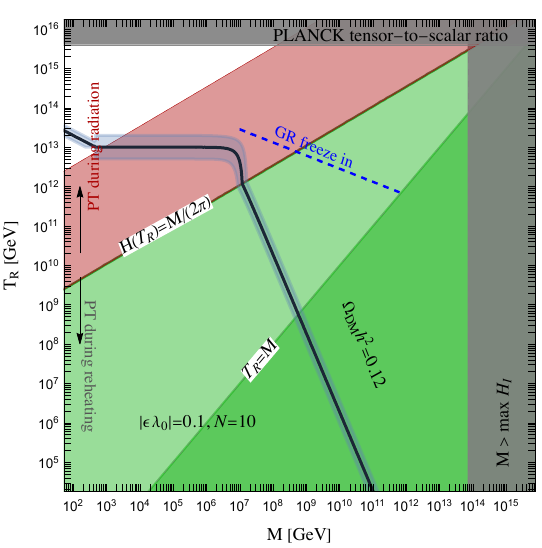}
 \caption{    \label{fig:plotRS} \small \it Parameter space of Dilaton DM. The DM abundance is reproduced along the black solid line. The darker green region corresponds to $T_R\leq \Lambda$ where the gravitational freeze-in (blue line) becomes inefficient (see eq.~\eqref{eq:GR-0}). The lighter green region corresponds to $H_R\leq \Lambda$ where the phase transition happens during reheating. The red region is highlighted to show where the contribution from phase transition generated in radiation domination dominate over the gravitational freeze-in. }
\end{figure}

\subsection{Dilaton Dark Matter}
So far we discussed the pure CFT limit where the system has no mass gap. 
As well known a gapped system can be realized introducing marginal deformations (operators of dimensions $\Delta=4+\epsilon$ with $\epsilon \ll 1$) 
that grow in the infrared eventually creating a mass gap \cite{Rattazzi:2000hs}. This is the analog of confinement in gauge theories.
The mechanism is entirely analogous to the Coleman-Weinberg mechanism  of dimensional transmutation in weakly coupled massless theories \cite{Coleman:1973jx}.

In the 5D picture bulk fields of mass $m$ correspond to operators of dimension $\Delta = 2+ \sqrt{4+m^2/k_5^2}$ so that marginal deformations
are mapped into 5D scalar fields with mass $|m|\ll k_5$. Through the Goldberger-Wise mechanism the effective potential induced 
by the scalars stabilizes the radius of the extra-dimensions to a finite value \cite{Goldberger:1999uk}. This can be described by 
the dilaton effective action \cite{Agashe:2019lhy},
\be\label{lagrangian-normalized}
\mathscr{L}= \frac {N^2}{16\pi^2}[ (\partial \varphi)^2 -  \hat V(\varphi)]+V_0\,,\quad
\hat V(\varphi)= \lambda_0 \varphi^4 \bigg[ 1 -\frac{4}{4+\epsilon} \bigg(\frac{\varphi}{\Lambda}\bigg)^\epsilon\bigg]+{\cal O}(\lambda_0^2)\,,
\ee
where $V_0=-\frac {N^2}{64\pi^2} \frac {\epsilon}{1+\epsilon/4} \lambda_0 \Lambda^4$ is the vacuum energy. We have introduced $\epsilon=\Delta-4\approx m^2/(4k_5^2)$.
Expanding around the minimum $\varphi=\Lambda$ we find
\begin{equation}
\Delta V =\frac{N^2}{64\pi^2}|\epsilon \lambda_0| \Lambda^4 \,, ~~~~~~~ M^2 =-2 \epsilon\lambda_0 \Lambda^2
\end{equation}
so that the dilaton mass is suppressed compared to the dynamical scale $4\pi/N \Lambda$.

In the 5D picture the IR scale is related to the position of the IR brane $\Lambda\sim k_5 \exp[-k_5 y_{\rm IR}]$ that is stabilized by the Goldberger-Wise fields.
This leads to a phase transition: if $y_{\rm IR}> y_{\rm hor}$ in (\ref{eq:hor}) the brane is replaced by the AdS horizon and the CFT is in the unbroken phase in de-Sitter background. 
In the opposite regime the 4D curvature becomes negligible. This is somewhat similar to the finite temperature phase transition
of Randall-Sundrum models \cite{Creminelli:2001th}. In the present case however the phase transition, as for glueballs, would be non thermal.

Deformations of the CFT by marginally relevant operators break conformal invariance similarly to 
$\beta$ functions in gauge theories. From the discussion in section \ref{sec:gauge-theories} it follows that the production during inflation will then
be suppressed by $(4-\Delta)^2 \approx m^4/k_5^4$. 

As in section \ref{sec:PT} we assume that the phase transition completes when $H=M/(2\pi)$ and that the energy density redshifts as matter afterwards.
The abundance of dilaton DM is found,
\be
\frac{\rho_{\rm dilaton}}{s}\bigg|_{\rm PT}=\frac{N^2}{64\pi^2}\frac{1}{|\epsilon \lambda_0|}\, \mathrm{min}\left[0.5  \frac{M^{5/2}}{\Mpl^{3/2}}, \frac{T_R M^2}{\Mpl^2}\right]
\ee

This contribution can be enhanced at small $\epsilon$ leading to smaller dilaton masses than the one of the glueball case. For reference, in figure \ref{fig:plotRS} we consider the case of $N=10$ and $|\epsilon\lambda_0|=0.1$.

Let us now discuss the stability. In the case of glueballs the critical mass is in tension with cosmological stability of DM.
For the dilaton the situation is more promising because the decay rate is suppressed by the mass. 
The dilaton decay to gravitons arises from higher dimension operators such as $\varphi R_{\mu\nu\rho\sigma}^2/\Lambda$. This leads to a decay 
rate $\Gamma \sim 1/(8\pi) M^3 \Lambda^2/(\Mpl^4)$ that is suppressed by $16\pi^2/N^2 \Lambda^2/M^2$ compared to the glueball estimates.
A full computation could be performed in the 5D RS scenario.

\section{Conclusions}\label{sec:conclusions}
Particle production in dark sectors is often discussed at the level of free field theories and leads to few viable scenarios.
In this work we have shown that secluded dark sectors with non trivial dynamics can be populated through 
a phase transition where the energy of the false vacuum is converted into massive stable particles. 
This genesis of the dark sector from the vacuum opens new avenues for dark sector model building.
The presence of interactions is the crucial ingredient of the framework that can in principle give rise to different DM
phenomenology.

We focused in particular on approximately Weyl invariant dark sectors with dynamical mass scale $M$.
This is rather generic being automatically realized for example in confining gauge theories and more general 
in interacting Conformal Field Theories. Assuming the scale of inflation to be large compared to $M$ we
have shown in general that inflationary production is strongly suppressed being determined by the
$\beta$ functions of the theory. This in turn avoids  strong constraints from isocurvature perturbations.
On the other hand inflation offers a mechanism to prepare the initial state
in a false vacuum. The energy will then be released during the evolution of the universe.
Since the control parameter is the Hubble parameter we argued that this happens when $H\sim M$ with an energy $M^4$.

We found that this contribution can dominate the universal gravitational freeze-in  and ordinary particle production in 
the expanding universe, if the reheating temperature is not maximal. The general prediction is that DM
is heavy in the range $10^8$ GeV or larger raising the question of cosmological stability of DM.

The dynamics considered in this paper is non-standard and would deserve further study.
While for the scalar we can explicitly follow the evolution, the non-thermal phase transition of Yang-Mills theories or
Randall-Sundrum scenarios due to the rapid expansion of the universe would be interesting on its own.
In this work we did not consider the possibility of a phase transition during inflation
that might lead to different scenarios with lighter DM. We hope to return to these questions in future work.

{\small
\subsubsection*{Acknowledgements}
This work is supported by MIUR grants PRIN 2017FMJFMW.  AT also thanks the MITP in Mainz for hospitality during the completion of this work.
MR would like to thank Savas Dimopoulos, Mina Arvanitaki and Giovanni Villadoro for discussions.}

\appendix

\section{Conformally coupled scalar}\label{app:A}
In this appendix we review the derivation of the particle production for a massive real scalar field $X(\eta,\vec x)$ conformally coupled to the metric, during radiation domination. We write the quantum real scalar field as
\be\label{X-scalar}
X(\eta, \vec x)=\int \frac{d^3k}{(2\pi)^3} X(\eta,\vec k) e^{-i \vec x \cdot \vec k} =\frac{1}{a} \int \frac{d^3k}{(2\pi)^3} \bigg(v_k(\eta) b_{\vec{k}} e^{-i \vec x \cdot \vec k} + h.c\bigg)
\ee
where the quantum operators annihilates the Bunch Davies vacuum at the beginning of inflation and the mode functions are $v_k(\eta)=\exp(-i k\eta)/\sqrt{2k}$ at $\eta =-\infty$ in the asymptotic past.  The equation for the mode function is 
\be
v_k''(\eta)+\omega_k^2(\eta) v_k(\eta)=0\,, \quad \omega_k^2(\eta)\equiv k^2+M^2 a^2(\eta)\,.
\label{eq:mode}
\ee
The computation of particle production requires the knowledge of the mode function $v$ in the asymptotic future in radiation domination, from which we can extract the Bogoliubov coefficient $|\beta_k(\eta)|^2$. This is essential in the computation of the energy density (and number density) at $\eta\to \infty$
\be
a^4 \frac{d \rho}{d\log k}= \frac{k^3}{2\pi^2} \lim_{\eta\to \infty} \left[\frac{|\partial_\eta v_k|^2}{2}+\frac{\omega_k^2|v_k|^2}{2}-\frac{\omega_k}{2}\right]= \frac{k^3}{2\pi^2} \lim_{\eta \to \infty} \omega_k |\beta_k|^2\,.
\ee

For a conformally coupled scalar with mass $H_I>M$, the maximal production happens in radiation when $H$ drops below $M$ and $k/a\approx M$, while production during inflation is very negligible. We assume that at the onset of radiation the relevant modes are not distorted from the Bunch-Davies conditions, and they satisfy eq. (\ref{eq:mode})
with boundary condition,
\be\label{eq:in-radiation}
v_k(\eta_e)=\exp(-i k\eta_e)/\sqrt{2k}\,,\quad v_k'(\eta_e)=- i \sqrt{\frac k 2}\exp(-i k\eta_e)
\ee
The full solution can be written in general as
\be
v_k(\eta)= \alpha_k(\eta) \exp(-i \int^{\eta} \omega_k d\eta)/\sqrt{2\omega_k(\eta)} + \beta_k(\eta) \exp(i \int^{\eta} \omega_k d\eta)/\sqrt{2\omega_k(\eta)}
\ee
where $\alpha_k(\eta_e)=1$ and $\beta_k(\eta_e)=0$. Deep into radiation domination the scale factor can be written as $a(\eta)\propto \eta$. For a sharp transition to radiation at the end of inflation at $\eta_e=0$ we can write $a(\eta)=a_R (1+ a_R H_R \eta)\approx a_R^2 H_R \eta$. The equation then becomes simply
\be
v_k''(\eta) + (k^2 +M^2 a_R^4 H_R^2\eta^2)v_k(\eta)=0
\ee
which is accurate for modes $k\ll k_{\rm max}\equiv a_R H_R$. The above equation is solved by a parabolic cylinder function, solution of the following 
\be
D''(w) + (\nu +\frac{1}{2}  -\frac{w^2}{4}) D(w)=0,\quad D_{\nu}(w)=\left\{\begin{array}{cc} w^\nu e^{-w^2/4} & w\to \infty \\ \frac{2^{\nu/2}\sqrt{\pi}}{\Gamma[\frac12-\frac{\nu}{2}]}& w\to 0 \end{array}\right.\,.
\ee
From the form of the equation, by symmetry we can construct a second solution via $w\to \pm iw$ as well as $\nu\to-1-\nu$. Solution of the equation with $+w^2/4$ can be constructed straightforwardly. In our case, deep into radiation domination, the mode function is then
\be
v_k(\eta)=c_1 D_\nu (\sqrt{2 M H_R} a_R \eta e^{i\pi/4}) + c_2 D_{\nu^*}(-i \sqrt{2 M H_R} a_R \eta e^{i\pi/4})\,, \quad \nu= - i \frac{k^2}{2a_R^2 H_R M}- \frac{1}{2}\,.
\ee
Up to irrelevant complex phases, the Bogoliubov coefficient is 
\be
\beta_k(\infty) = c_2 \,(2a_R^2 H_R M)^{\frac14} \exp(\frac{k^2 \pi}{8a_R^2 H_R M})\,
\ee
where the coefficient $c_2$ is matched to the initial value in \eqref{eq:in-radiation}. Noticing that $\omega_k \approx M a(\eta)$ at late times, we have
\be
a^3 \frac{d\rho}{d\log k}= \frac{k^3}{2\pi^2} M |\beta_k(\infty)|^2\,.
\ee 
From the expansion of the parabolic cylinder functions we get the full expression at $\eta=\infty$,
\be
|\beta_k(\infty)|^2=\frac{e^{-\frac{3 z}{4}} \left(e^z+1\right) }{4\sqrt{\pi z}} \left(\frac{z}{4\pi}  \bigg|\Gamma \left(\frac14 + i\frac{z}{4\pi}\right)\bigg|^2+  \bigg|\Gamma \left(\frac{3}{4} +i \frac{z}{4\pi}\right)\bigg|^2\right)-\frac12\,,\quad  z\equiv \frac{\pi k^2}{a_R^2 H_R M}\,.
\ee
By direct inspection we extract the following behaviour 
\be
|\beta_k (\infty)|^2 = \left\{\begin{array}{cc} \displaystyle \frac{\Gamma[3/4]^2}{2\pi}\, a_R \frac{\sqrt{H_R M}}{k}
&\mathrm{small}~\, k \\
\displaystyle a_R^8\frac{M^4 H_R^4}{64 k^8}& \mathrm{large}~\, k\end{array}\right.\,.
\ee
The above formula completes our calculation and allows us to derive the energy density as 
\be
a^3 \frac{d\rho}{d\log k}= \frac{k^3}{2\pi^2} M \left\{\begin{array}{cc} \displaystyle \frac{\Gamma[3/4]^2}{2\pi}\, a_R \frac{\sqrt{H_R M}}{k}
&\mathrm{small}~\, k \\
\displaystyle a_R^8\frac{M^4 H_R^4}{64 k^8}& \mathrm{large}~\, k\end{array}\right.\,.
\ee
From the measure $k^3$ we see that in the IR $(k\ll M)$ the power spectrum goes like $k^2$, while it is strongly suppressed for $k\gg M$ as $k^{-5}$. 
From the full expression for $|\beta_k|^2$ at infinity we can numerically integrate the power spectrum on all $k$. In terms of the dimensionless variable $z$, the integral of the power gives $\mathcal{C}=\int dz/(2z) \times (z^{3/2}|\beta|^2)\approx 0.165$. The peak of the power spectrum is located at $k_{\rm peak}\approx 0.63 a_R \sqrt{M H_R}\approx 10^{24} \, \mathrm{Mpc}^{-1}\, \sqrt{M/\Mpl}$. From the integral, we can finally write the energy density
\be
\frac{\rho}{s}=\frac{1}{s}\mathcal{C}\frac{a_R^3}{a^3} \frac{M^{5/2} H_R^{3/2}}{2\pi^{7/2}}=0.0012499 \left(\frac{100}{g(T_R)}\right)^{1/4} \mathcal{C}\, M\left(\frac{M}{\Mpl}\right)^{\frac32}\,.
\ee

\section{Weyl Fermion}
The discussion can be directly generalized to massive fermions.
We consider here a 2-component massive Weyl fermion $\psi$ (see \cite{Ema:2019yrd} for Majorana notation), with the following Lagrangian
\be
\mathscr{L}=i \bar \psi \bar\sigma^\mu (\partial_\mu + \omega_\mu)\psi -\frac{M}{2}(\psi\psi+\bar\psi\bar\psi)\,.
\ee
Here $\omega_\mu$ is the spin connection to be computed in the Friedmann-Robertson-Walker background and $M$ is real. With the field redefinition $\psi\equiv \chi/a^{3/2}$ in an expanding background the Dirac equation takes the form
\begin{equation}\label{eq:Dirac}
i \bar{\sigma}^\mu \partial_\mu \chi = M a(\eta) \bar{\chi}\,.
\end{equation}
Upon quantization the field can be expanded into creation and annihilations operators as
\be
\chi(\eta,\vec x) = \sum_{h=\pm}\int \frac{d^3k}{(2\pi)^3}\bigg[ v_{\vec k, h}(\eta)\xi_{\vec k, h}e^{+i \vec k \cdot \vec x} a_{\vec k,h} + v^*_{\vec k, -h}(\eta)[\epsilon\xi_{\vec k, h}^*]e^{-i \vec k \cdot \vec x} a^\dag_{\vec k,h}]\,.
\ee
Here we have introduced eigenstates of the helicity operator, $\vec \sigma \cdot \vec k\xi_{\vec k, h}= h |\vec k|\xi_{\vec k, h}$. They are normalized to unity and satisfy $\epsilon \xi_{\vec k, h}^*=h \xi_{\vec k, -h}$ up to a phase. 

The wave-functions $v_{\vec k,\pm}$ are solutions to the classical equations of motion \eqref{eq:Dirac}
\begin{eqnarray}\label{eq:dirac2}
&&i v_{\vec k, -}' - k v_{\vec k, -} =  -M a(\eta) v_{\vec k, +}\\
&&i v_{\vec k, +}' + k v_{\vec k, +} =-  M a(\eta) v_{\vec k, -}\,
\end{eqnarray}
These can be be normalized as
\be\label{eq:norm}
|v_{\vec k,-}(\eta)|^2 + |v_{\vec k, +}(\eta)|^2=1, \quad \quad \forall k,\eta
\ee
that implies $\{ a_{\vec k,h}\,,a^\dag_{\vec k',h'}\}=(2\pi)^3 \delta^{(3)}(\vec k -\vec k') \delta_{hh'}$. At  early times, the field is effectively massless and the positive frequency wave is given by
\be
v_{\vec k, -}(\eta\to-\infty)=e^{-i k \eta}, \quad v_{\vec k,+}(\eta\to-\infty)=0\,,\quad \quad \forall k\,,
\ee
as a massless Weyl fermion in flat space.

At late times it is easy to check that the solutions are of the form
\be\label{eq:matching-v}
v_{\vec k, \pm}(\eta) =\frac{\alpha_{\vec k}}{\sqrt{2}} e^{-i \int^\eta M a\, d\eta'} \pm \frac{\beta_{\vec k}}{\sqrt{2}} e^{+i \int^\eta M a\, d\eta'}\,,
\ee
From the normalization  \eqref{eq:norm} one finds,
\be
|\alpha_{\vec k}|^2+|\beta_{\vec k}|^2=1\,,
\ee
as required for the coefficients of the Bogoliubov transformation.

In order to determine the solution,  it is convenient to define the following two combinations of mode functions
\be
\chi^\pm_{\vec k}\equiv v_{\vec k, +}\pm v_{\vec k, -}\,,
\ee
that enjoy the property that only $\chi^{\pm}_{\vec k}$ has a positive/negative frequency waves in the future, and solve the second order differential equations
\be
(\partial_\eta^2 + k^2 + a^2 M^2 \mp i (aM)') \chi^\pm_{\vec k} =0\,.
\ee
The boundary condition in the far past are inherited by $v_{\vec k, \pm}$ and read
\be
\chi^{-}_{\vec k}(\eta\to-\infty)=-e^{-i k\tau}\,,\quad\quad \chi^{+}_{\vec k}(\eta\to-\infty)=e^{-i k\tau}\,.
\label{eq:appbc}
\ee
Using (\ref{eq:matching-v}) $\beta_{\vec k}$ can be extracted from the coefficient  of the negative energy component of $\chi^{-}_{\vec k}$ for $\eta\to \infty$. The solution to the above second order differential equation is again found in terms of parabolic cylinder functions, as for the conformally coupled scalar.  During a phase of radiation dominance, where $a(\eta)=a_R^ 2H_R \eta$ we have
\be
\chi^-_{\vec k}(\eta)=c_1 D_{-1-i \frac{k^2}{2a_R^2 H_R M}}(a_R\sqrt{2MH_R}\eta e^{-i\pi/4}) + c_2 D_{i \frac{k^2}{2a_R^2 H_RM}}(a_R\sqrt{2MH_R}\eta e^{-i\frac{3}{4}\pi})\,,
\ee
where $c_{1,2}$ are fixed by the boundary conditions (\ref{eq:appbc}). The Bogoliubov coefficient $\beta_{\vec k}$  is given by,
\be
\beta_{\vec k}=\frac{c_2}{\sqrt{2}} \exp(-\frac{3k^2\pi}{8a_R^2 H_R M})\,.
\ee
Defining again $z\equiv \pi k^2/(a_R^2 H_R M)$, one finds,
\be
|\beta_k(\infty)|^2=\frac{e^{-\frac{5 z}{4}} \left| \left(-1+e^{z}\right) \Gamma \left(1-\frac{i z}{2 \pi}\right)
   \left(\sqrt[4]{-1}  \Gamma \left(\frac{i z}{4 \pi}\right)-2 \sqrt{\frac{\pi}z} \Gamma \left(\frac{i z}{4
   \pi}+\frac{1}{2}\right)\right)\right| ^2}{32 \pi ^2}\,,
\ee
whose expansion is
\be
|\beta_k(\infty)|^2 = \left\{\begin{array}{cc} \displaystyle 1/2
&\mathrm{small}~\, k \\
\displaystyle a_R^4\frac{M^2 H_R^2}{16 k^4}& \mathrm{large}~\, k\end{array}\right.\,.
\ee
The abundance in eq. (\ref{eq:quantumrad}) is now determined by the integral
\be
\mathcal{C}=\int dz/(2z) \times (z^{3/2}|\beta_k|^2)\approx 0.6\,,
\ee
that is 4 times larger than the scalar per each chirality. We thus find that 
the Weyl fermion reproduces the cosmological abundance for $M=2.6 \cdot 10^8$ GeV,
if produced during radiation domination.

\pagestyle{plain}
\bibliographystyle{jhep}
\small
\bibliography{biblio}

\end{document}